\documentclass[10pt]{letter}
\usepackage{amsmath}
\usepackage{amssymb}
\usepackage{multicol}
\usepackage[dvips]{graphicx}
\DeclareFontFamily{U}{msb}{}
\DeclareFontShape{U}{msb}{m}{n}{ <5> <6> <7> <8> <9> gen * msbm
        <10> <10.95> <12> <14.4> <17.28> <20.74> <24.88> msbm10}{}
\DeclareSymbolFont{AMSb}{U}{msb}{m}{n}
\DeclareMathSymbol{\realset}{\mathalpha}{AMSb}{"52}
\newcommand{\mbi}{\ensuremath{\mathbf{i}}}
\newcommand{\mbj}{\ensuremath{\mathbf{j}}}

\begin{document}

\begin{center}
\Large 
{\bf A central partition of molecular conformational space. II. 
Embedding $3D$ structures.}
\end{center}

\begin{center}
{\large J. Gabarro-Arpa} \\
\ \ \ Ecole Normale Sup\'erieure de Cachan, LBPA, C.N.R.S. UMR 8113 \\
\ \ \ 61, Avenue du Pr\'esident Wilson, 94235 Cachan cedex, France \ \ \ -
\ \ \  jga@infobiogen.fr \\
\end{center}

\begin{multicols}{2}
\hspace*{4mm} {\it Abstract---}
{\bf A combinatorial model of molecular conformational space
that was previously developped [1], had the drawback that structures 
could not be properly embedded beacause it lacked explicit rotational 
symmetry. The problem can be circumvented by sorting the elementary
$3D$ components of a molecular system into a finite set 
of classes that can be separately embedded. 
This also opens up the possibility of encoding 
the dynamical states into a graph structure.}

\hspace*{4mm} {\it Keywords---}
{\bf Molecular conformational space, hyperplane arrangement, face lattice,
     molecular dynamics } 

\begin{center}
{\scshape I. Introduction}
\end{center}
\hspace*{4mm} In a previous paper [1] it was presented a combinatorial model 
of molecular confomational space (thereafter refered as CS), it was shown that 
it could be described with a fair degree of accuracy by a central arrangement 
of hyperplanes\footnote{the term {\bf central} means that all the hyperplanes 
pass through the origin.}
that partitions the space into a set of cells. The arrangement was defined
such that, for a molecule of $N$ atoms, the 3-dimensional ($3D$) conformations 
in a cell all have the same dominance sign vector: 
for a given vector $p$ $\epsilon$ $\realset^N$ there is an associated dominance
sign vector $\mathcal{D}(p)=(d_{12},d_{13}, ... ,d_{p-2,p},d_{p-1,p})$
whose components are defined as follows

\hspace*{12mm} $d_{ij} =  \left\{\begin{array}{rr}
 + & \mbox{ $p_{i} < p_{j} $ } \\
 0 & \mbox{ $p_{i} = p_{j} $ } \\
 - & \mbox{ $p_{i} > p_{j} $ } \\
\end{array}\right.$ \ $ 1 \leq i < j \leq N$ \ \ \ \ \ \ (1)

There is a set of three dominance sign vectors per $3D$ conformation one 
for each coordinate: the partition is actually a product 
of three partitions [1]. \\
\hspace*{4mm} A central concept for the combinatorial study of an hyperplane 
arrangement is the face lattice poset [2]: the cells in the induced 
decomposition of $\realset^{3N-3}$ ordered by inclusion. It is this 
hierarchical structure that enables us to manage the sheer complexity 
of CS since with the simple codes (1) we can encompass from broad regions 
down to single cells. \\
\hspace*{4mm} The model takes into account two basic symmetries of CS: \\
\hspace*{4mm} 1) The translation symmetry: for a molecule with $N$ atoms 
                 the model is build in a $(3N-3)$-dimensional subspace, 
                 since for each $x$, $y$ or $z$ coordinates the dimension 
                 parallel to the vector $(1,1,1,...)$ contains conformations 
                 that are obtained by translation along the axis. \\
\hspace*{4mm} 2) The scaling symmetry: the points lying on a half-line 
                 starting at the origin result from multiplying 
                 the coordinates of a given $3D$ conformation by an arbitrary 
                 positive factor. It reflects the fact that the unit length 
                 in our system can be arbitrarily defined. \\
\hspace*{4mm} The model however fails to incorporate the all important rotation symmetry,
this is due to the fact that combinatorial approaches like ours apply mostly
to linear systems. 
This greatly complicates the embbeding of $3D$ conformations. \\
\hspace*{4mm} To circumvent this problem, the approach we explore 
in the present communication is how conformations can be embedded in CS 
starting from its elementary building blocks, and the most elementary 
component $3D$ structure is a simplex\footnote{A three-dimensional polytope 
with four vertices.}.
Many structural patterns in molecular systems can be decomposed 
into simplexes [3]. \\
\hspace*{4mm} As we shall see below embedding in CS just a simplex 
is not simple, but this approach leads us to study a set of combinatorial
structures that offer, beyond the embedding problem, the interesting 
possibility of encoding the dynamical states of a molecular system.

\begin{center}
{\scshape II. Embedding a simplex}
\vskip 2mm
\includegraphics{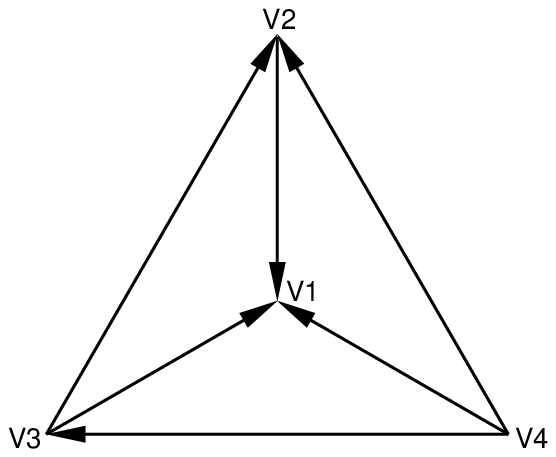}
\end{center}
{\footnotesize Fig. 1. Graph of a simplex, with the vectors along the edges 
oriented as to make the graph acyclic. If we assume that $v_1$ lies above the 
plane of the figure, it corresponds to a right-handed simplex. }
\vskip 2mm
\hspace*{4mm} From the simplex of Fig. 1 we define the following set 
of vectors and their associated central planes:
\vskip 2mm
\hspace*{10mm} $ e_{\mbi\mbj}=v_{\mbi}-v_{\mbj} , \ \
                 \mathcal{E}_{\mbi\mbj}^0(x)=\{x \ \epsilon \ \realset^3 : 
 e_{\mbi\mbj}.x=0\} $ \ \ \ \ (2)
\vskip 2mm
for $1 \leq \mbi < \mbj \leq 4$. These six planes generate a partition of 
$3D$ space into 24 cells [4] (see Fig. 2): each plane divides
the space into positive and negative hemispaces and a zero space in between
\vskip 2mm
\hspace*{17mm} $\mathcal{E}_{\mbi\mbj}^+(x)=\{x \ \epsilon \ \realset^3 :
               e_{\mbi\mbj}.x > 0\}$ \ \ and \\
\hspace*{17mm} $\mathcal{E}_{\mbi\mbj}^-(x)=\{x \ \epsilon \ \realset^3 :
e_{\mbi\mbj}.x < 0\}$ 
\ \ \ \ \ \ \ \ \ \ \ \ \ \ \ \ \ (3)
\vskip 2mm
a $3D$ cell results from the intersection of six hemispaces, thus it 
can be unambiguously characterized by the signs 
of these six hemispaces (Fig. 2). \\
\hspace*{4mm} It is easy to see that the dominance sign vectors 
for an arbitrary $3D$-reference system centered at the origin 
can be obtained from this partition: consider for instance the $z$-axis and 
suppose that $z.e_{12} > 0$, this means that $z$ is in a cell where 
the $e_{12}$ component of the sign vector is $+$, which in turn implies 
that $v_{1z} > v_{2z}$. \\
\hspace*{4mm} Thus the dominance sign vector for each coordinate
will be the sign vector of the cell that contains its positive semi-axis.
\vskip 2mm
\begin{center}
\includegraphics{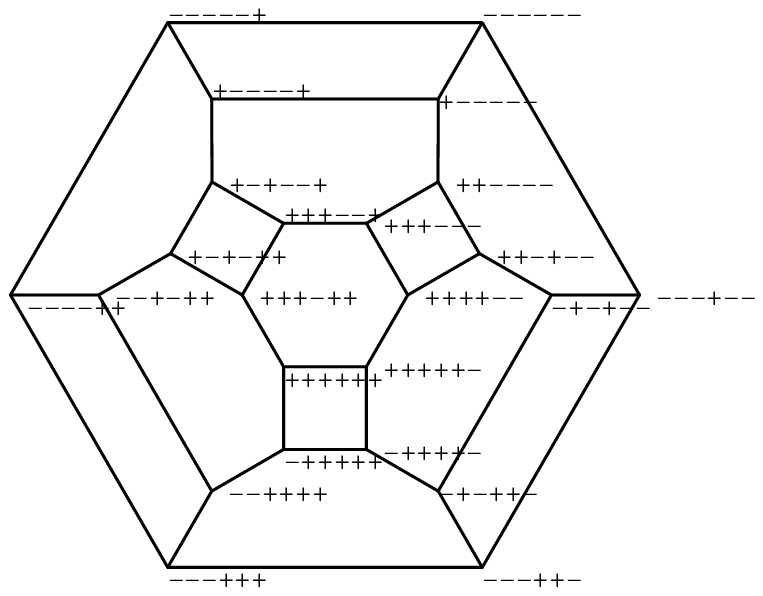}
\end{center}
\vskip 2mm
{\footnotesize Fig. 2. Tope graph of the partition (2), each node corresponds
to a $3D$ cell and the edges represent the planes separating the cells. 
For each node the corresponding sign vector is annotated on the right.}

\hspace*{4mm} Lower dimensional cells occur for vectors that lie in one 
or more of the planes $\mathcal{E}_{\mbi\mbj}$, in that case the corresponding 
components of the sign vector are zero. \\
\hspace*{4mm} The $1$-dimensional ($1D$) cells are rays starting at the origin
and running parallel to the vectors

\hspace*{1.5mm} $f_{123}=e_{12} \wedge e_{23}, f_{124}=e_{12} \wedge e_{24}, 
                 f_{134}=e_{13} \wedge e_{34},$  \\
\hspace*{1.5mm} $f_{234}=e_{23} \wedge e_{34}, f_{12,34}=e_{12} \wedge e_{34},
                 f_{13,24}=e_{13} \wedge e_{24},$ \\
\hspace*{1.5mm} $f_{14,23}=e_{14} \wedge e_{23}$ 
\ \ \ \ \ \ \ \ \ \ \ \ \ \ \ \ \ \ \ \ \ \ \ \
\ \ \ \ \ \ \ \ \ \ \ \ \ \ \ \ \ \ \ \ \ \ \ \ (4)
\vskip 2mm
the first four are the vectors perpendicular to the faces of the simplex,
the last three are perpendicular to pairs of non-adjacent edges.
As in (2) they have a set of associated central planes
\vskip 2mm
\hspace*{9mm} $\mathcal{F}_{123}^0,   \mathcal{F}_{124}^0,
               \mathcal{F}_{134}^0,   \mathcal{F}_{234}^0,
               \mathcal{F}_{12,34}^0, \mathcal{F}_{13,24}^0, 
               \mathcal{F}_{14,23}^0$ \ \ \ \ \ (5)
\hspace*{20mm} $ \mathcal{F}_{\alpha}^0(x)=\{x \ \epsilon \ \realset^3 : 
                           f_{\alpha}.x=0\} $

The corresponding sign vectors are the rows in the matrix below

\hspace*{14mm} $ \begin{matrix}
          & e_{12} & e_{13} & e_{14} & e_{23} & e_{24} & e_{34} \\
f_{123}   &  0     &  0     &  +     &  0     &  +     &  +     \\
f_{124}   &  0     &  -     &  0     &  -     &  0     &  +     \\
f_{134}   &  +     &  0     &  0     &  -     &  -     &  0     \\
f_{234}   &  +     &  +     &  +     &  0     &  0     &  0     \\
f_{12,34} &  0     &  -     &  -     &  -     &  -     &  0     \\
f_{13,24} &  +     &  0     &  +     &  -     &  0     &  +     \\
f_{14,23} &  -     &  -     &  0     &  0     &  +     &  +
  \end{matrix}$ 
\ \ \ \ \ \ \ \ (6)
\vskip 2mm
The zeros in the matrix correspond to the planes that intersect the 
corresponding $1D$ cell, this means that the cells encoded 
by $f_{123}$ and $f_{12,34}$, for instance, are sourrounded by six and four 
$3D$ cells respectively. \\
\hspace*{4mm} The zeros in the sign vector of lower dimensional
cells can be seen as a sort of wildcard: they match the sign vectors of all
the adjacent cells. The converse is also true: a sign vector from a $3D$
cell can be obtained by adding up the sign vectors from the adjacent 
lower dimensional cells [4]. As an example, for the lower left cell of Fig. 2 
we have
$(---+++) = (-00++0) + (--00++) + (---000) = 
-SIGN(f_{134}) + SIGN(f_{14,23}) -SIGN(f_{234})$.

\begin{center}
{\scshape III. Embedding a simplex}
\end{center}

\hspace*{4mm} There is still another set of sign vectors that will be most 
useful in characterizing the geometric properties of simplexes: these are
the signs of the scalar products of the vectors (2) and (4) between them. \\
\hspace*{4mm} Let us assume that we have a particular right-handed simplex 
whose set of signs is
\vskip 2mm
\hspace*{18mm} $ \begin{matrix} 
          & e_{13} & e_{14} & e_{23} & e_{24} & e_{34} \\
e_{12}    &  -     &  -     &  -     &  -     &  -     \\
e_{13}    &        &  +     &  +     &  +     &  -     \\
e_{14}    &        &        &  +     &  +     &  +     \\
e_{23}    &        &        &        &  +     &  -     \\
e_{24}    &        &        &        &        &  +
\end{matrix}$ 
\ \ \ \ \ \ \ \ \ \ \ \ \ (7a)

\hspace*{5mm} $ \begin{matrix} 
          & f_{124} & f_{134} & f_{234} & f_{12,34} & f_{13,24} & f_{14,23} \\
f_{123}   &  +      &  +      &  +      &  -        &  +        &  +        \\
f_{124}   &         &  -      &  -      &  +        &  +        &  +        \\
f_{134}   &         &         &  +      &  -        &  +        &  -        \\
f_{234}   &         &         &         &  -        &  +        &  -        \\
f_{12,34} &         &         &         &           &  -        &  +        \\
f_{13,24} &         &         &         &           &           &  -
\end{matrix}$ 
 (7b)

\vskip 2mm
\hspace*{4mm} The set of sign vectors (7a) refers mostly to the angles between 
adjacent edges while (7b) are mostly related to dihedral angles between 
contiguous faces: $+$, $0$ and $-$ are for acute, right and obtuse angles 
respectively. Thus (7) gives us a rough outline of the geometry of a 
simplex, and allows a classification of simplexes. \\
\hspace*{4mm} Next we are going to proceed to embed our simplex in CS for 
the particular case where the $z$-axis runs paralell to $f_{123}$. 
The reason for this special choice will be explained below. \\
\vskip 2mm
\includegraphics{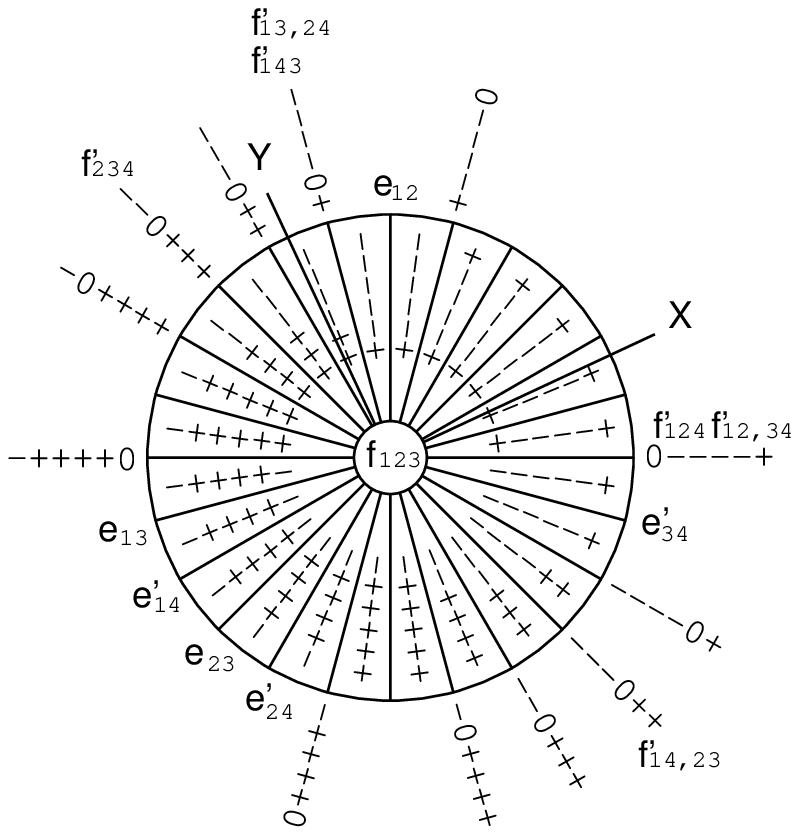}
\vskip 2mm
{\footnotesize Fig. 3. View from above of $\mathcal{F}_{123}^0$,
the plane appears divided in a number of sectors, separating sectors are rays 
running along vectors whose denomination appears at the outer extreme, vector 
names bearing a $'$ are projections: 
$x'=x-f_{123} (x.f_{123})/\|f_{123}\|$. Rays orthogonal to vectors
$e_{\mbi\mbj}'$ always correspond to intersections with the planes 
$\mathcal{E}_{\mbi\mbj}^0$. In the inner layer each sector harbors 
its corresponding sign vector, in the outer layer are the sign vectors
corresponding to the $1D$ cells. 
Sign vectors should be read in the outward direction.
The $x$ and $y$ axis are depicted in a random position.}

\hspace*{4mm} From Fig. 3 we can see that $\mathcal{F}_{123}^0$ 
has been partitioned into 24 cells, which are delimited by lines along the 
projections of vectors $e_{\mbi\mbj}$ and from the intersections 
with planes $\mathcal{E}_{\mbi\mbj}^0$ (2): thus each projected 
$e_{\mbi\mbj}'$ vector is perpendicular to the intersection of the 
corresponding $\mathcal{E}_{\mbi\mbj}^0$ and it divides the
plane into two signed moities as in (2) and (3), so the sign vector 
associated with each cell in Fig. 3 becomes obvious, by construction they are
the dominance sign vectors associated to a coordinate axis 
that is in the cell.
\hspace*{4mm} Also by construction, the sixth cell after/before the one 
under consideration is orthogonal to it. Thus by rotating the $x$ and $y$ axis 
around $z$ in Fig. 3 we can scan the complete set of dominance sign
vectors that arise for this particular situation. Before proceeding further
let us explain how Fig. 3 can be obtained from (6) and (7). \\
\hspace*{4mm} By construction, see Fig. 1 and (4), $e_{12}$, $e_{13}$ 
and $e_{23}$ are in circular counter-clockwise order, with the last 
two vectors in the negative hemispaces of $e_{12}$ (7a) and $f_{124}$ (6). 
For the remaining projections: \\
\hspace*{4mm} $e_{14}'$)  $e_{14}$ lies above $\mathcal{F}_{123}^0$ (6), 
                          by (7b) the same is true of $f_{134}$ 
                          and $f_{14,23}$, as $e_{14}$ is contained 
                          in the plane perpendicular to these vectors 
                          $e_{14}'$ will be located inside the sector 
                          determined by $e_{13}$ and $e_{23}$. \\
\hspace*{4mm} $e_{24}'$)  $e_{24}$ is above $\mathcal{F}_{123}^0$ (6) 
                          as well as $f_{124}$ and $f_{234}$ (7b), using 
                          the same argument as above $e_{14}$ will be located 
                          inside the sector determined by -$e_{12}$ 
                          and $e_{23}$. \\
\hspace*{4mm} $e_{34}'$)  $e_{34}$ lies in the $- - - +$ sector relative 
                          to $e_{12}$, $e_{13}$ $e_{23}$ and $e_{24}$ (7a), 
                          this squeezes $e_{34}'$ between $\mathcal{E}_{12}^-$ 
                          and $\mathcal{E}_{24}^-{'}$.

\begin{center}
{\scshape IV. The circular order of the projected vectors}
\end{center}
\begin{center}
\includegraphics{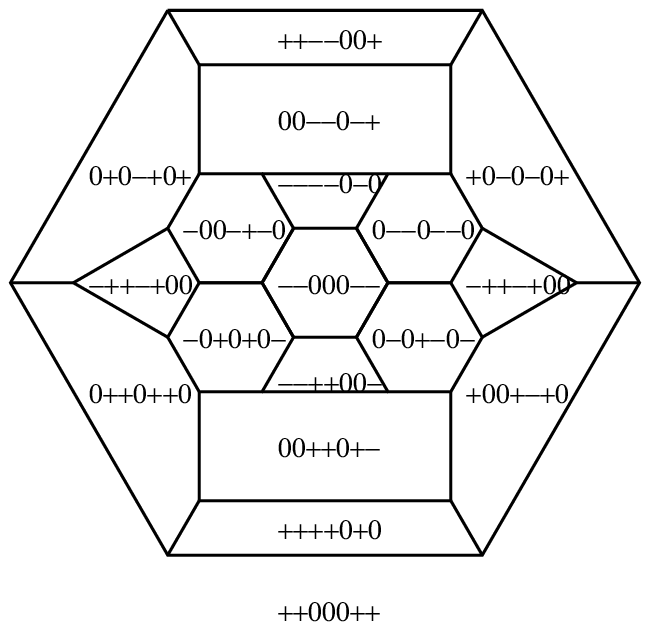}
\end{center}
{\footnotesize Fig. 4. Tope graph of the partition (4) showing the sign 
 vectors of $1D$ cells. As in Fig. 2 the graph is planar, sign vectors of 
 $3D$ cells can be obtained by adding the sign vectors around each node.
 Notice that the columns in (6), or their centrally symmetric sign vectors,
 all correspond to sign vectors above.}

\hspace*{4mm} The circular order of the projected vectors in Fig. 3 can 
be obtained from the signs of the first row in (7b): one can see from Fig. 3 
that the shortest path between $e_{12}$ and $e_{34}'$ runs clokwise, 
while the shortest path from $e_{23}$ to $e_{24}'$ and $e_{34}'$ runs 
counterclockwise. This is simply due to the fact that $e_{23}$, $e_{24}$ 
and $e_{34}$, for instance, are contained in $\mathcal{F}_{234}^0$ and the
angle between $e_{23}$ and $e_{34}$ can not exceed $2 \pi$, thus
their circular order in the projection depends on wether $f_{234}$ lies
above or below $\mathcal{F}_{123}^0$. \\
\hspace*{4mm} To obtain the sign vector encoding the circular order in 
the plane perpendicular to a vector in general position, it suffices to
look wether the given vector is in the positive or negative hemispace 
relative to the planes (5).
Thus the central arrangement generated by (4) partitions 
the space in 32 cells that are represented by the tope graph of Fig.4. \\
\hspace*{4mm} This settles the problem of the circular ordering of the 
projected vectors which is completely determined by (6). \\
\hspace*{4mm} Last but not least, it is indispensable for the correct 
simultaneous allocation of the $x$ and $y$ dominance sign vectors, 
to determine the relative positions in the circular ordering between 
the $e_{\mbi\mbj}'$s and the intersections 
of the $\mathcal{E}_{\mbi\mbj}^0$s. \\
\hspace*{4mm} As this is not a linear problem in some cases it can only 
be partially resolved by (7). Ambiguities can arise when building 
a projection, for instance: if in the example of Fig. 3 $e_{24}'$ and the 
intersection of $\mathcal{E}_{34}^0$ both fell in the same sector. 
In that case we would have to split the diagram into two alternative ones.  \\

\begin{center}
{\scshape V. The enumeration of minimal vectors }
\end{center}

\hspace*{4mm} In the diagram from Fig. 3 the dominance sign vector 
associated with the $\{x, y, z\}$ reference frame is 
\begin{center}
$((+----+),(++----),(00+0++))$,
\end{center}
one can notice that it is squeezed between 

\hspace*{6mm} $((+----+),(++0---),(00+0++))$ and \\
\hspace*{6mm} $((+----+),(+0----),(00+0++))$.

\hspace*{4mm} The importance of this diagram is that 
it enumerates all the lower dimensional $2D$ cells associated with the sign 
vector $(00+0++)$, and since the rows of (6) are all the $1D$ cells 
in the partition constructing a diagram like the one in Fig. 3 for every row 
in (6) allows us to enumerate all the minimal vectors in our system 
(those bearing a maximum number of zeros), 
all other sign vectors being combinations of them.

\begin{center}
{\scshape VI. The general embedding problem }
\end{center}

\hspace*{4mm} In molecular dynamics simulations atoms are represented by
pointlike structures surrounded by a force field, thus any four atoms 
in a molecular structure can form a simplex.
If an order relation has been defined between the atoms of the system, 
then vectors (2) and (5) can be defined too for every simplex with
the node numbers of Fig. 1 representing the order of the atoms. \\
\hspace*{4mm} Some of the vectors (2) and (5) are shared between simplexes 
through common edges and faces, as a consequence orienting a simplex restricts 
the range of available orientations in the other simplexes.
Embedding a $3D$ conformation in CS can be done with this simple algorithm: \\
\hspace*{4mm} 1) take a set of connected simplexes\footnote{
                 $\mathbf{S_a}$ and $\mathbf{S_b}$ are connected if there
                 exists a sequence of simplexes $\{\mathbf{S_\mbi}\}$ 
                 $1 \leq \mbi \leq N$, with $\mathbf{S_1}=\mathbf{S_a}$ and
                 $\mathbf{S_N}=\mathbf{S_b}$,
                 such that for $1 \leq i < N$ $\mathbf{S_\mbi}$ 
                 and $\mathbf{S_{\mbi+1}}$ are adjacent.}
                 such that every pair of atoms in the structure is at least 
                 in one simplex, \\
\hspace*{4mm} 2) choose a simplex with a non empty set of available 
                 sign vectors, otherwise terminate the procedure, \\
\hspace*{4mm} 3) select one orientation and restrict the available 
                 orientations in the other simplexes to the ones compatible 
                 with this choice. Repeat step 2. \\

\begin{center}
{\scshape VII. Conclusion}
\end{center}

\hspace*{4mm} The two main results of this communication are \\
\hspace*{4mm} 1) simplexes can be put into a number of discrete classes, 
                 not taking into account handedness we have: 258 and 816 
                 sets for (7a) and (7b) respectively, with both combined 
                 we have a total of 3936 classes. \\
\hspace*{4mm} 2) these classes are related to cells in CS, thus relating 
                 the binary sets (7) in a molecular conformation 
                 to $3D$ coordinates. \\
\hspace*{4mm} Embedding just one $3D$ conformation is not an interesting issue,
what really matters is embedding the volume occupied by a molecular system. \\ 
\hspace*{4mm} Beyond the embedding problem the results above offer 
the possibility of building a structure encoding the dynamical states 
of a molecule. This can be seen by analyzing the dynamical activity 
of all simplexes in a typical molecular dynamics simulation like 
the one studied in [5], we find that \\
\hspace*{4mm} 1) 90\% of the simplexes evolve within less than 20 classes, \\
\hspace*{4mm} 2) 0.4\% remain in a single class for the duration 
                 of the simulation, form a connected set and comprise 95\% 
                 of the dominance relations (1), \\
\hspace*{4mm} 3) the most dynamically active simplex spans a range of 171 
                 classes, slightly less than 2\% of the total. \\
\hspace*{4mm} A connected set of simplexes can obviously be represented 
by a graph, where each node can be split into a number of classes (7): 
its dynamical states. 
The connectivity between the sets (7) is an issue than has not been 
explored in this paper, but its determination will allow the connexion
between dynamical states in adjacent nodes thus generating the derived graph
of the molecular system dynamical states. \\
\hspace*{4mm} This graph is a subject for further research.

\begin{center}
{\scshape References}
\end{center}
\begin{itemize}

\item[[1]] J. Gabarro-Arpa, "A central partition of molecular conformational
           space. I. Basic structures" {\it Comp. Biol. and Chem., } 
           27, pp. 153-159, 2003.

\item[[2]] J. Folkman, J. Lawrence, "Oriented matroids" {\it J. Combinatorial
           Theory, } ser. B, 25, pp. 199-236, 1978.

\item[[3]] J.F. Sadoc, R. Mosseri, "Geometrical frustration". Cambridge ,UK:
           Cambridge University Press, 1999.
 
\item[[4]] A. Bjorner, M. las Vergnas, B. Sturmfels, N. White, "Oriented 
           Matroids". Cambridge, UK: Cambridge University Press, 1993, ch. 1-2,
           pp. 1-64.
 
\item[[5]] C. Laboulais, M. Ouali, M. Le Bret, J. Gabarro-Arpa, "Hamming
           distance geometry of a protein conformational space: application
           to the clustering of a 4-ns molecular dynamics trajectory of the
           HIV-1 integrase catalytic core" 
           {\it Proteins: Struct. Funct. Genet. } 47, pp. 169-179, 2002.

\end{itemize}

\end{multicols}

\end{document}